\documentclass[12pt]{amsart}
\usepackage{amsmath, amsthm, latexsym, amssymb, amsfonts, epsfig, epsf, xcolor, hyperref}
\usepackage{mathtools}
\usepackage{bm}

\newenvironment{pf}{\proof[\proofname]}{\endproof}
\theoremstyle{plain}
\newtheorem{Th}{Theorem}[section]
\newtheorem{Cor}[Th]{Corollary}

\newtheorem{Prop}[Th]{Proposition}
\newtheorem{Lemma}[Th]{Lemma}
\numberwithin{equation}{section} \theoremstyle{definition}
\newtheorem{Rem}[Th]{Remark}

\newtheorem{Ex}[Th]{Example}

\newtheorem{Def}[Th]{Definition}

\newcommand{\cal}[1]{\mathcal{#1}}
\newcommand{\F}{\mathbb F}

\newcommand{\Z}{\mathbb Z}
\newcommand{\R}{\mathbb R}

\newcommand{\cL}{\cal L}

\newcommand{\cS}{\cal S}

\newcommand{\ev}{\operatorname{ev}}
\newcommand{\Res}{\operatorname{Res}}

\newcommand{\spn}{\operatorname{Span}}

\newcommand{\supp}{\operatorname{supp}}

\newcommand{\Ker}{\operatorname{Ker}}

\newcommand{\rr}[1]{Remark~\ref{R:#1}}
\newcommand{\rex}[1]{Example~\ref{ex:#1}}

\begin{document}


\title[Monomial-Cartesian codes with app. to LCD, quantum and LRC codes]{Monomial-Cartesian codes and their duals, with applications to LCD codes, quantum codes, and locally recoverable codes}
\author{Hiram H. L\'opez}
\address[Hiram H. L\'opez]{Department of Mathematics\\ Cleveland State University\\ Cleveland, OH USA}
\email{h.lopezvaldez@csuohio.edu}
\author{Gretchen L. Matthews}
\address[Gretchen L. Matthews]{Department of Mathematics\\ Virginia Tech\\ Blacksburg, VA USA}
\email{gmatthews@vt.edu}
\thanks{The second author was supported by NSF DMS-1855136.}
\author{Ivan Soprunov}
\address[Ivan Soprunov]{Department of Mathematics\\ Cleveland State University\\ Cleveland, OH USA}
\email{i.soprunov@csuohio.edu}
\thanks{}

\begin{abstract}
A monomial-Cartesian code is an evaluation code defined by evaluating a set of monomials
over a Cartesian product. It is a generalization of some families of codes in the literature, for instance toric codes, affine Cartesian codes
and $J$-affine variety codes. In this work we use the vanishing ideal of the
Cartesian product to give a description of the dual of a monomial-Cartesian code. Then we use such description of the dual to prove the existence of
quantum error correcting codes and MDS quantum error correcting codes.
Finally we show that the direct product of monomial-Cartesian codes is a locally recoverable code with $t$-availability if at least $t$ of the components are locally recoverable codes.
 
\end{abstract}

\maketitle

\section{Introduction}
Let $K=\mathbb{F}_q$ be a finite field with 
$q$ elements and $R=K[x_1,\ldots,x_m]$ the polynomial ring over $K$ in $m$ variables. We write $K^*=K \setminus \{ 0 \}$ for the multiplicative group of $K$. Given a lattice point $\bm{a}\in\Z_{\geq 0}^m$ we use $\bm{x}^{\bm{a}}$ to denote the corresponding monomial in $R$, i.e.
$\bm{x}^{\bm{a}}=x_1^{a_1}\cdots x_m^{a_m}$ for $\bm{a}=(a_1,\dots,a_m)$.
Given a positive integer $\ell,$ we define $\left[\ell\right]:=\left\{1,\ldots,\ell\right\}.$

A { monomial-Cartesian} code is defined as follows. Fix non-empty subsets $S_1,\ldots,S_m$ of $K$. 
Define their {\it Cartesian product} as
\[\cS:=S_1\times\cdots\times S_m\subseteq K^{m}.\]
Furthermore, let $A\subset\Z_{\geq 0}^m$ be a finite lattice set and $ \cL(A)$ the subspace of polynomials of $R$ that are $K$-linear
combinations of monomials with exponents in $A$:
$$ \cL(A)=\spn_K\{\bm{x}^{\bm{a}} : \bm{a}\in A\}\subset R.$$
Fix a linear order of the points in
$\cS=\{\bm{s}_1,\ldots,\bm{s}_n\},$ $\bm{s}_1 \prec \cdots \prec \bm{s}_n$.
This defines the {\it evaluation map}

\[
\begin{array}{lclll}
{\ev_{\cS}}\colon & \cL(A)\ & \to & K^{|\mathcal{\cS}|}\\
&f &\mapsto &  \left( f(\bm{s}_1),\ldots,f(\bm{s}_n)\right).
\end{array}
\]
In what follows, $n_i:=|S_i|$, the cardinality of $S_i$ for $i\in[m].$
From now on, we assume that $A\subset \{0,\ldots, n_1-1\}\times \cdots \times \{0,\ldots, n_m-1\},$
that is the degree of each $f\in \cL(A)$ in $x_i$ is less than
$|S_i|$. In this case the evaluation map $\ev_{\cS}$ is injective (see the proof of Proposition~\ref{19.04.30}). 

\begin{Def} Let $\cS\subset K^m$ and $A\subset\Z_{\geq 0}^m$ be as above. The image $\ev_{\cS}(\cL(A))\subset K^{|\cS|}$ is called the {\it monomial-Cartesian code} associated with $\cS$ and $A$. We denote it by $C(\cS,A)$. By an abuse of notation, if $\bm{a}\in A$ then $C(\cS,\bm{a})$ denotes the code
$C(\cS,\{\bm{a}\}).$
\end{Def}

The monomial-Cartesian code has the following parameters (Proposition~\ref{19.04.30}).
Its length and dimension
are given by $n=|\cS|$ and $k=\dim_K C(\cS,A)=|A|$, respectively.
Recall that the {\it minimum
weight\/} of a code $C$  is given by 
\[
\delta(C)=\min\{|\supp(c)| : 0\neq c\in C\},
\]
where $\supp(c)$ denotes the support of $c$, that is the set of all non-zero entries of $c$. Unlike the case of the length and the dimension, in general, there is no explicit formula for $\delta C(\cS,A)$ in terms of $\cS$ and $A$. For toric codes, some explicit formulas appear in \cite{Sop} and non-trivial bounds appear in \cite{Sop2} when $m=2$. In the following proposition we mention a simple formula for the direct product of monomial-Cartesian codes. We consider them in Section~\ref{19.07.2}. 
\begin{Prop}\label{19.07.30}
Let $C(\cS_1,A)$ and $C(\cS_2,B)$ be monomial-Cartesian codes and consider their direct product $C(\cS_1\times \cS_2,A\times B)$. Then $$\delta\left(C(\cS_1\times \cS_2,A\times B)\right)=\delta\left(C(\cS_1,A)\right)\delta\left(C(\cS_2,B)\right).$$
\end{Prop}
Previous result can be proven doing a slight
modification to the proof of \cite[Theorem 2.1]{Sop} or because \cite[Theorem 3 c)]{Wei}.
There is also an inductive lower bound for $\delta\left(C(\cS,A)\right)$ in terms minimum weights of 
monomial-Cartesian codes corresponding to projections and fibers of $A$ along coordinate subspaces. It 
is stated in \cite[Theorem 4.1]{Sop3} in the case of generalized toric codes, but the statement and the 
proof can be easily adapted to arbitrary monomial-Cartesian codes.

%

The {\it dual} of the code $C$ is defined by
\[C^{\perp} = \{ \bm{w}\in K^n : \bm{w}\cdot\bm{c}=0 \text{ for all } \bm{c}\in C \}, \]
where $\bm{w}\cdot\bm{c}$ represents the {\it Euclidean inner product}. The code $C$ is called a
{\it linear complementary dual} ({\it LCD}) \cite{JMassey} if $C\cap C^{\perp}=\{ \bm{0} \},$ and is called a {\it self-orthogonal} code if $C^{\perp} \subset C.$ In \cite{Carlet_equiv}, Carlet, Mesnager, Tang, Qi, and Pellikaan show that any linear code over $\F_q$ with $q > 3$ is equivalent to an LCD code; even so, explicit constructions can be elusive. In this paper, we provide a characterization for monomial-Cartesian codes which are LCD, thus providing explicit constructions of LCD codes.

Instances of monomial-Cartesian codes for particular families of lattice sets $A$ and Cartesian products $\cS$
have been extensively studied in the literature. For example, a Reed-Muller code of order $r$ in the sense of 
\cite[p.~37]{tsfasman} is the {monomial-Cartesian} code $C(K^m,A_r),$
where $A_r=\{(a_1,\ldots,a_m)\in\Z^m_{\geq 0} : a_1+\dots+a_m\leq r\}$. Note that in this case $\cL(A_r)=R_{\leq r}$, the set of all polynomials of degree at most $r$. 

Another example of a monomial-Cartesian code is a {\it toric code} $C(\left(K^*\right)^m,A_P),$ where
$A_P=P\cap\Z^m$ is the set of lattice points of a convex lattice polytope $P\subset\R^m$
and $\left(K^*\right)^m$ is the Cartesian product with $S_1=\cdots=S_m=K^*.$ Good references for toric codes are \cite{Han1, joyner-decoding, Sop}.

An {\it affine Cartesian code of order $r$} is a {monomial-Cartesian} code $C(\cS,A_{r}),$ where $A_r$ is as above and
$\cS$ is an arbitrary Cartesian set. This family of codes appeared first time
in \cite{Geil} and then independently in \cite{lopez-villa}.
In \cite{Geil}, the authors study the basic parameters of Cartesian codes, they determine
optimal weights for the case when $A_r$ is the Cartesian product of two sets, and then
present two list decoding algorithms. In \cite{lopez-villa} the authors study
the vanishing ideal $I(\mathcal{S})$. Using commutative algebra tools such as regularity, degree, and Hilbert function,
the authors determine
the basic parameters of Cartesian codes in terms of the size of the components
of the Cartesian product.
In \cite{carvalho}, the author shows some results on higher
Hamming weights of Cartesian codes and gives a different proof for
the minimum distance using the concepts of
Gr\"obner basis and footprint of an ideal. In \cite{carvalho2} the authors
find several values for the second least weight of codewords, also known as the
next-to-minimal Hamming weight. In \cite{BeelenDatta} the authors find the
generalized Hamming weights and the dual of Cartesian codes. In \cite{lopez-manga-matt}
the authors study the dual of a generalized Cartesian product and the property of being
LCD, { i.e.}, when the code and the dual have zero intersection.

Let $\cS\subset K^m$ and $A\subset\Z_{\geq 0}^m$ be as above.
In this work we are interested in the properties and applications of the monomial-Cartesian code $C(\cS,A)$. In Section~\ref{19.07.01}
we give a nice description of the dual of the code $C(\cS,A)$ in terms of the complement of the set $A$ and the vanishing ideal of the
set of points $\cS.$ Our main theorem generalizes some results of \cite{BrME, GHRuano, GOlav} and \cite{DRua}, where the dual of toric codes,
$J$-affine variety codes and generalized toric codes are studied. The representation for the dual gives rise to a Goppa 
representation for $C(\cS,A)$, which may open the path for an efficiently decoding algorithm, because such representation is the key
to decode the well-known Reed-Solomon codes. It is important to remark that there are decoding algorithms in the literature that can be used to decode
particular cases of monomial-Cartesian codes, but the complexity is not as good as the one for the Reed-Solomon codes. For instance, the decoding algorithm 
developed by \cite{FaSh} depends of finding a Gr\"obner basis for each received codeword, and it would decode monomial-Cartesian codes on the case when
$\cS$ is arbitrary and $A\subset\Z_{\geq 0}^m$ are the smallest elements for a fixed monomial order in $Z_{\geq 0}^m.$
Excellent references about how to decode linear codes using Gr\"obner basis are \cite{BulPel1,BulPel2,BulPel3} and \cite{BulPel4}.

The monomial-Cartesian code construction provides the flexibility needed for some applications, such as that of quantum error-correcting codes and locally recoverable codes. Quantum codes support resilience of quantum information by correcting bit and phase flip errors in qudits, quantum digits, which is fundamental to fault-tolerant quantum computation. While the goal of quantum codes is similar to that of linear codes, new techniques are needed for their construction due to the inability to duplicate quantum information. Even so, there is a link between quantum codes and classical linear codes, due to independent work of Calderbank and Shor \cite{CS} and Steane \cite{Steane}. Indeed, the CSS construction uses linear codes which contain their duals to construct quantum codes.  A family of codes called $J$-affine variety codes were introduced and studied in \cite{GHRuano} and \cite{GOlav}, respectively. This family of codes can be seen
as monomial-Cartesian codes $C(\cS,A)$ with the condition that $n_i-1$ divides $q-1.$ Inspired by those works, where the authors
use $J$-affine variety codes to prove the existence of quantum error correcting codes we use monomial-Cartesian codes in Section~\ref{19.07.29}
to prove the existence of quantum error correcting codes with certain parameters. An $[[n,k,d]]_q$ quantum code satisfies the quantum Singleton bound \cite{KKKK}
\[k\leq n-2d+2.\] If $k= n-2d+2$ then the quantum code is called {\it quantum maximum-distance-separable} (MDS) code. We obtain quantum MDS codes from monomial-Cartesian codes, making use of knowledge of the dual.

The idea of a locally recoverable code is that every coordinate depends of a few other coordinates. By ``depends" we mean that if one of the coordinates
is erased, then that coordinate can be recovered using some coordinates. Of course, it is desirable that ``some" is small. The concept of
$t$-availability means that for any coordinate there are $t$ pair disjoint subsets of a few coordinates each in such a way that the each subset can be used
to recover such coordinate. Traditionally, for locality and availability it is assumed that the received coordinates are correct, but it may happens in practice that
the received coordinates that are not erased contain also errors. Previous situation with errors gives rise to the codes known as locally recoverable codes with 
local error detection, which was introduced recently in \cite{Mun}.
Section~\ref{19.07.2} we study local properties for direct product of monomial-Cartesian codes.

More information about basic theory for coding theory can be found in \cite{huf-pless,MacWilliams-Sloane,van-lint}.
More constructions of evaluation codes can be seen in
\cite{carvalho3,carvalho4,GRT,renteria-tapia-ca2}. Excellent references for theory of vanishing ideals and its properties are \cite{CLO1,Eisen,harris,monalg}.

\section{Dual of Monomial-Cartesian codes}\label{19.07.01}
Denote the variables $x_1,\ldots,x_m$ by $\bm{x}.$
An important characteristic for monomial-Cartesian codes and evaluation codes in general
is the fact that we can use commutative algebra methods to
study them. The kernel of the evaluation map
${\rm ev}_\cal{S}$ is precisely $\cal{L}\left(A\right)\cap I(\mathcal{S})$, where $I(\mathcal{S})$
is the {\it vanishing ideal\/} of $\mathcal{S}$ consisting of 
all polynomials of $R$ that vanish on ${\mathcal{S}}$.
Thus,  algebraic properties of $R/\left(\cal{L}\left(A\right)\cap I(\mathcal{S})\right)$
are related to the basic parameters of $C(\mathcal{S},A)$.
For each $i\in[m],$ define the polynomial
\begin{equation}\label{03-24-18}
L_i(x_i):=\prod_{\substack{s_{j}\in S_i}}\left(x_i-s_{j}\right).
\end{equation}
The vanishing ideal of the Cartesian product $\mathcal{S}$ is given by
$
I(\mathcal{S})=
\left(L_1(x_1),\ldots,L_m(x_m)\right),
$
\cite[Lemma 2.3]{lopez-villa}. Moreover, 
let $\prec$ be the {\it graded-lexicographic order} on the set of monomials of $R.$
This order is defined in the following way:
$x_1^{a_1}\cdots x_m^{a_m}\prec x_1^{b_1}\cdots x_m^{b_m}$ if and only if
$\sum_{i=1}^m a_i<\sum_{i=1}^m b_i$ or $\sum_{i=1}^m a_i=\sum_{i=1}^m b_i$
and the leftmost nonzero entry in $(b_1-a_1,\ldots,b_m-a_m)$ is positive.
From now on we fix the order $\prec$.
Then, according to \cite[Proposition 4]{CLO1}, $\left\{
L_1(x_1),\ldots,L_m(x_m)
\right\}$
is a Gr\"obner basis of $I(\mathcal{S})$, relative to the order $\prec$.
\begin{Prop}\label{19.04.30}
The dimension and the length of the monomial-Cartesian code $C(\mathcal{S},A)$
are given by $|A|$ and $|\cal{S}|,$ respectively.
\end{Prop}
\begin{pf} It is enough to show that  the evaluation map
${\rm ev}_\mathcal{S}\colon \cal{L}\left(A\right)\  \to  K^{|\mathcal{S}|}$
is injective. By above $\Ker({\rm ev}_\mathcal{S})=\cal{L}\left(A\right)\cap I(\mathcal{S})$. On one hand, by assumption
$\deg_{x_i}(f)<n_i$ for every $f\in \cL(A)$ and $i\in[m]$. On the other hand, $I(\mathcal{S})$ has a Gr\"obner basis 
$\left\{L_1(x_1),\ldots,L_m(x_m)\right\}$ with $\deg_{x_i}(L_i)=n_i$ for each $i\in[m]$. Therefore,
$\cal{L}\left(A\right)\cap I(\mathcal{S})$ is trivial.
\end{pf}
\begin{Def}
For $\bm{s}=\left(s_1,\ldots,s_m\right)\in \mathcal{S}$ and
$f\in R,$ define the {\it residue} of $f$ at $\bm{s}$ as
\begin{equation}\label{03-23-18}
\Res_{\bm{s}}f=f(\bm{s})
\left(
\prod_{i=1}^m \prod_{\substack{s_{i}^\prime\in S_i\setminus\{s_i\}}}\left(s_i-s_i^{\prime}\right)
\right)^{-1}.
\end{equation}
For simplicity, we introduce the following notation for the residues vector
\[\Res_{\cal S}f=
\left(\Res_{\bm{s}_1}f,\ldots, \Res_{\bm{s}_n}f \right).\]
\end{Def}

\begin{Rem}\label{R:residue-map}
Note that $\Res_{\cal S}:R\to K^{|\cS|}$ is a linear map which is injective on the subspace of polynomials $f$ satisfying $\deg_{x_i}(f)<n_i$.
This follows from the definition of the residue and the proof of Proposition~\ref{19.04.30}.
\end{Rem}

By  \cite[Theorem 5.7]{BeelenDatta} or \cite[Theorem 2.3]{lopez-manga-matt}, the dual of the monomial-Cartesian code
$C(\mathcal{S},\bm{0})\subset K^{|\mathcal{S}|},$
 where $\bm{0}$ is the zero vector in $\mathbb{Z}^m_{\geq 0},$
is given by
\[
C(\mathcal{S},\bm{0})^{\perp}=
\spn_K \Big\{ \Res_{\cal S}f
: \deg(f)< \sum_{i = 1}^m (n_i - 1),\deg_{x_i}(f)<n_i \Big\}.
\]
This implies
\begin{equation}\label{19.05.02}
\sum_{i=1}^n \Res_{\bm{s}_i}f=0, \text{ for }
f\in R \text{ with } \deg(f)< \sum_{i = 1}^m (n_i - 1) \text{ and } \deg_{x_i}(f)<n_i.
\end{equation}
By the division algorithm there are polynomials $q_{i,j}$ and $r_{i,j}$ in
$K[x_i]$  for $i\in [m]$, such that
\begin{equation}\label{19.05.04}
L_i=x_i^{j}q_{i,j-1}+r_{i,j-1},
\end{equation}
and $\deg(r_{i,j-1})< j$. For every 
$\bm{b}=\left(b_1,\ldots,b_m \right)$ in $\mathbb{Z}^m_{\ge 0}$ 
define the polynomial
\begin{equation}\label{19.05.01}
Q_{\bm{b}}(\bm{x}):=\prod_{i=1}^m q_{i,b_i}(x_i).
\end{equation}
These polynomials $Q_{\bm{b}}\in R$ help to describe the dual of a monomial-Cartesian code.

\begin{Lemma}\label{19.05.05} Let $B=\{0,\ldots,n_1-1\}\times\cdots\times\{0,\ldots,n_m-1\}.$
For any $\bm{a}\in B$, the set 
$\left\{\Res_{\cal S} Q_{\bm{b}} : \bm{b}\in B, \bm{b} \neq \bm{a}\right\}$ forms a basis for the dual $C(\mathcal{S},\bm{a})^\perp$ of the
monomial-Cartesian code $C(\mathcal{S},\bm{a})$.
\end{Lemma}
\begin{pf}
By definition, $\deg_{x_i}(Q_{\bm{b}})=n_i-(b_i+1).$ This implies that the $Q_{\bm{b}}$ for $\bm{b}\in B$ have pairwise distinct multidegrees
(with respect to the graded-lexicographic order). Thus the set
$\{Q_{\bm{b}} : \bm{b}\in B, \bm{b} \neq \bm{a}\}$ is linearly independent. Furthermore, by \rr{residue-map}, its image under the residue map
$\left\{
\Res_{\cal S} Q_{\bm{b}}
: \bm{b}\in B, \bm{b} \neq \bm{a}\right\}
$ spans a subspace of dimension $\sum_{i = 1}^m (n_i - 1)-1=\dim C(\bm{a},\mathcal{S})^{\perp}$.

Now we check the inner product. 
Let $\overline{f}$ denote the normal form of $f$ with respect to the Gr\"obner basis 
$\left\{L_1(x_1),\ldots,L_m(x_m)\right\}$. Note that $\Res_{\bm{s}_i}f=\Res_{\bm{s}_i}\overline{f}$ for any $i\in[n]$ and $f\in R$. Therefore,
$$
\left( \bm{s}_1^{\bm{a}},\ldots, \bm{s}_n^{\bm{a}} \right)\cdot\Res_{\cal S}Q_{\bm{b}}=
\sum _{i=1}^n \bm{s}_i^{\bm{a}} \Res_{\bm{s}_i}Q_{\bm{b}}=
\sum _{i=1}^n \Res_{\bm{s}_i} \bm{x}^{\bm{a}} Q_{\bm{b}}=
\sum _{i=1}^n \Res_{\bm{s}_i} \overline{\bm{x}^{\bm{a}} Q_{\bm{b}}}.
$$
It remained to show that $\sum _{i=1}^n \Res_{\bm{s}_i} \overline{\bm{x}^{\bm{a}} Q_{\bm{b}}}=0$. For this we check the conditions
in Equation~(\ref{19.05.02}). We have
\begin{equation}\label{19.05.03}
\deg_{x_i} \big(\overline{\bm{x}^{\bm{a}}Q_{\bm{b}}(\bm{x})}\big)\leq
\left\{\hspace{-1mm}
\begin{array}{cll}
 a_i+n_i-(b_i+1)&\mbox{ if } a_i\leq b_i,\\
a_i-1&\mbox{ if } a_i > b_i.
\end{array}
\right.
\end{equation}
Indeed, the first inequality is clear. For the second one, when $a_i > b_i$ the devision algorithm and Equation~\ref{19.05.04} provide
$$\deg_{x_i} \big(\overline{\bm{x}^{\bm{a}}Q_{\bm{b}}(\bm{x})}\big)=
\deg_{x_i}\big(\overline{x_i^{a_i}q_{i,b_i}(x_i)}\big)=
\deg_{x_i} \big(x_i^{a_i-(b_i+1)}r_{i,b_i}(x_i)\big)< a_i.$$
Now, since $\bm{a}\neq \bm{b}$, there is $j\in[m]$ such that $a_j\neq b_j.$ Then (\ref{19.05.03}) implies
$$\deg_{x_j}\big(\overline{\bm{x}^{\bm{a}}Q_{\bm{b}}(\bm{x})}\big)< n_j-1.$$
Also, (\ref{19.05.03}) provides $\deg_{x_i}\big(\overline{\bm{x}^{\bm{a}}Q_{\bm{b}}(\bm{x})}\big)< n_i$ for all $i\in[m]\setminus\{j\}$.
Therefore, both conditions of (\ref{19.05.02}) are satisfied which shows that $\sum _{i=1}^n \Res_{\bm{s}_i} \overline{\bm{x}^{\bm{a}} Q_{\bm{b}}}=0$.
\end{pf}

\begin{Ex}\label{ex:1} 
Let $K=\mathbb{F}_7$ and assume $\cS=\{1,3,4,5\}\subset K.$ On this case
$L_1(x_1)=(x_1-1)(x_1-3)(x_1-4)(x_1-5)$ and
\begin{center}
\begin{tabular}{ll}
$L_1(x_1)=x_1\underbrace{\left(x_1^3 + x_1^2 + 3x_1 + 5\right)}_{q_{0}(x_1)}+\underbrace{4}_{r_{0}(x_1)},$
&
$L_1(x_1)=x_1^2\underbrace{\left(x_1^2 + x_1 + 3\right)}_{q_{1}(x_1)}+\underbrace{5x_1 + 4}_{r_{1}(x_1)},$
\\\\
$L_1(x_1)=x_1^3\underbrace{\left(x_1+1\right)}_{q_{2}(x_1)}+\underbrace{3x_1^2 + 5x_1 + 4}_{r_{2}(x_1)},$
&
$L_1(x_1)=x_1^4\underbrace{\left(1\right)}_{q_{3}(x_1)}+\underbrace{x^3 + 3x_1^2 + 5x_1 + 4}_{r_{3}(x_1)}.$
\end{tabular}
\end{center}
Then we have the following duals
of $C(\mathcal{S},a)$ for $a\in A$.
\begin{center}
\begin{tabular}{ll}
$C(\mathcal{S},0)^{\perp}=\text{Span}_K \left\{
\Res_{\cal S}q_1,\Res_{\cal S}q_2,\Res_{\cal S}q_3\right\},$
&$C(\mathcal{S},1)^{\perp}=\text{Span}_K \left\{
\Res_{\cal S}q_0,\Res_{\cal S}q_2,\Res_{\cal S}q_3\right\},$\\
$C(\mathcal{S},2)^{\perp}=\text{Span}_K \left\{
\Res_{\cal S}q_0,\Res_{\cal S}q_1,\Res_{\cal S}q_3\right\},$
&
$C(\mathcal{S},3)^{\perp}=\text{Span}_K \left\{
\Res_{\cal S}q_0,\Res_{\cal S}q_1,\Res_{\cal S}q_2\right\}.$
\end{tabular}
\end{center}
\end{Ex}
\begin{Ex} Let $K=\mathbb{F}_7$.
Consider the following Cartesian set: $\cS=\{0,2,3\}\times\{0,1,3,5,6\}\subset K^2.$
On this case $L_1(x_1)=x_1(x_1-2)(x_1-3)$ and
$L_2(x_2)=x_2(x_2-1)(x_2-3)(x_2-5)(x_2-6)$. Then we have 
\begin{center}
\begin{tabular}{ll}
$L_1(x_1)=x_1\underbrace{\left(x_1^2+2x_1+6\right)}_{q_{1,0}(x_1)}+\underbrace{0}_{r_{1,0}(x_1)},$
&
$L_2(x_2)=x_2\underbrace{\left(x_2^4+6x_2^3+x_2+6\right)}_{q_{2,0}(x_2)}+\underbrace{0}_{r_{2,0}(x_2)},$
\\\\
$L_1(x_1)=x_1^2\underbrace{\left(x_1+2\right)}_{q_{1,1}(x_1)}+\underbrace{6x_1}_{r_{1,1}(x_1)},$
&
$L_2(x_2)=x_2^2\underbrace{\left(x_2^3+6x_2^2+1\right)}_{q_{2,1}(x_2)}+\underbrace{6x_2}_{r_{2,1}(x_2)},$
\\\\
$L_1(x_1)=x_1^3\underbrace{\left(1\right)}_{q_{1,2}(x_1)}+\underbrace{2x_1^2+6x_1}_{r_{1,2}(x_1)},$
&
$L_2(x_2)=x_2^3\underbrace{\left(x_2^2+6x_2\right)}_{q_{2,2}(x_2)}+\underbrace{x_2^2+6x_2}_{r_{2,2}(x_2)},$
\\\\
&
$L_2(x_2)=x_2^4\underbrace{\left(x_2+6\right)}_{q_{2,3}(x_2)}+\underbrace{x_2^2+6x_2}_{r_{2,3}(x_2)},$
\\\\
&
$L_2(x_2)=x_2^5\underbrace{\left(1\right)}_{q_{2,4}(x_2)}+\underbrace{6x_2^4 + x_2^2 + 6x_2}_{r_{2,4}(x_2)}.$
\end{tabular}
\end{center}
Then, the dual of 
$C(\mathcal{S},\bm{a})$ for $\bm{a}=(2,3)$
is given by
\[
C(\mathcal{S},(2,3))^{\perp}=\text{Span}_K \left\{
\Res_{\cal S}Q_{\bm{b}}
: b\in \{0,1,2\}\times\{0,1,2,3,4\}, b\neq (2,3)\right\}.
\]
In other words, we take the residue of all the products $q_{1,i}q_{2,j}$ except when $(i,j)$ is the given point $(2,3).$
\end{Ex}
\begin{Th}\label{19.06.26}
Let $\cS=S_1\times\cdots\times S_m\subseteq K^m$ and $B=\{0,\ldots, n_1-1\}\times \cdots \times \{0,\ldots, n_m-1\}\subset\Z^m$.
For any $A \subseteq B$, the set $\left\{\Res_{\cal S} Q_{\bm{b}}: \bm{b}\in B\setminus A\right\}$ forms a basis for the dual
$C(\mathcal{S},A)^{\perp}$ of the monomial-Cartesian code $C(\mathcal{S},A).$
\end{Th}
\begin{pf}
As for any two points $\bm{a}_1, \bm{a}_2 \in A$ we have that
$C(\mathcal{S},\{\bm{a}_1,\bm{a}_2\})^{\perp}=
C(\mathcal{S},\bm{a}_1)^{\perp} \cap C(\mathcal{S},\bm{a}_2)^{\perp},$
the result is a consequence of Lemma~\ref{19.05.05}.
\end{pf}
\begin{Ex}
Let $K=\mathbb{F}_7$ and assume $\cS=\{1,3,4,5\}\subset K$ as in \rex{1}. As before we have
$L_1(x_1)=(x_1-1)(x_1-3)(x_1-4)(x_1-5)$ and
\begin{center}
\begin{tabular}{ll}
$L_1(x_1)=x_1\underbrace{\left(x_1^3 + x_1^2 + 3x_1 + 5\right)}_{q_{0}(x_1)}+\underbrace{4}_{r_{0}(x_1)},$
&
$L_1(x_1)=x_1^2\underbrace{\left(x_1^2 + x_1 + 3\right)}_{q_{1}(x_1)}+\underbrace{5x_1 + 4}_{r_{1}(x_1)},$
\\\\
$L_1(x_1)=x_1^3\underbrace{\left(x_1+1\right)}_{q_{2}(x_1)}+\underbrace{3x_1^2 + 5x_1 + 4}_{r_{2}(x_1)},$
&
$L_1(x_1)=x_1^4\underbrace{\left(1\right)}_{q_{3}(x_1)}+\underbrace{x^3 + 3x_1^2 + 5x_1 + 4}_{r_{3}(x_1)}.$
\end{tabular}
\end{center}
Then we obtain the following dual codes. $C(\mathcal{S},\{2,3\})^{\perp}=\text{Span}_K \left\{
\Res_{\cal S}q_0,\Res_{\cal S}q_1\right\},$
\begin{center}
\begin{tabular}{ll}
$C(\mathcal{S},\{0,2\})^{\perp}=\text{Span}_K\left\{
\Res_{\cal S}q_1,\Res_{\cal S}q_3\right\}$
& and $C(\mathcal{S},\{1,2,3\})^{\perp}=\text{Span}_K \left\{
\Res_{\cal S}q_0\right\}.$
\end{tabular}
\end{center}
\end{Ex}

\begin{Ex} Let $K=\mathbb{F}_7$.
Consider the following Cartesian set: $\cS=\{0,2,3\}\times\{0,1,3,5,6\}\subset K^2.$
On this case $L_1(x_1)=x_1(x_1-2)(x_1-3)$ and
$L_2(x_2)=x_2(x_2-1)(x_2-3)(x_2-5)(x_2-6)$. We have 
\begin{center}
\begin{tabular}{ll}
$L_1(x_1)=x_1\underbrace{\left(x_1^2+2x_1+6\right)}_{q_{1,0}(x_1)}+\underbrace{0}_{r_{1,0}(x_1)},$
&
$L_2(x_2)=x_2\underbrace{\left(x_2^4+6x_2^3+x_2+6\right)}_{q_{2,0}(x_2)}+\underbrace{0}_{r_{2,0}(x_2)},$
\\\\
$L_1(x_1)=x_1^2\underbrace{\left(x_1+2\right)}_{q_{1,1}(x_1)}+\underbrace{6x_1}_{r_{1,1}(x_1)},$
&
$L_2(x_2)=x_2^2\underbrace{\left(x_2^3+6x_2^2+1\right)}_{q_{2,1}(x_2)}+\underbrace{6x_2}_{r_{2,1}(x_2)},$
\\\\
$L_1(x_1)=x_1^3\underbrace{\left(1\right)}_{q_{1,2}(x_1)}+\underbrace{2x_1^2+6x_1}_{r_{1,2}(x_1)},$
&
$L_2(x_2)=x_2^3\underbrace{\left(x_2^2+6x_2\right)}_{q_{2,2}(x_2)}+\underbrace{x_2^2+6x_2}_{r_{2,2}(x_2)},$
\\\\
&
$L_2(x_2)=x_2^4\underbrace{\left(x_2+6\right)}_{q_{2,3}(x_2)}+\underbrace{x_2^2+6x_2}_{r_{2,3}(x_2)},$
\\\\
&
$L_2(x_2)=x_2^5\underbrace{\left(1\right)}_{q_{2,4}(x_2)}+\underbrace{6x_2^4 + x_2^2 + 6x_2}_{r_{2,4}(x_2)}.$
\end{tabular}
\end{center}
Then, the dual of the code
$C(\mathcal{S},\{(0,1),(3,5)\})$ is given by
\[
C(\mathcal{S},\{(0,1),(3,5)\})^{\perp}=\text{Span}_K \left\{
\Res_{\cal S}Q_{\bm{b}}
: b\in \{0,1,2\}\times\{0,1,2,3,4\}, b\notin \{(0,1),(3,5)\}\right\}.
\]
In other words, we take the residue of all the products $q_{1,i}q_{2,j}$ except when $(i,j)$ is either $(0,1)$ or $(3,5).$
\end{Ex}

\section{Quantum error correcting codes}\label{19.07.29}
In this section, we give some applications of Monomial-Cartesian codes to quantum error correcting codes. Our main result
shows how to use monomial-Cartesian codes to find quantum error correction codes and MDS quantum error correction codes.
We continue using same
notation than previous sections, in particular
$\cS=S_1\times\cdots\times S_m\subseteq K^{m},$ $n_i:=|S_i|$,
$A\subset \{0,\ldots, n_1-1\}\times \cdots \times \{0,\ldots, n_m-1\},$
and $ \cL(A)=\spn_K\{\bm{x}^{\bm{a}} : \bm{a}\in A\}\subset R.$

First, we provide a slightly different representation for the dual of a monomial-Cartesian code.
\begin{Def}\label{19.07.24}
Let $F(\bm{x})$ be the unique element in $R$ such that $\deg_{x_i}F(\bm{x})<n_i$ and
$F(\bm{s})=
\left(
\prod_{i=1}^m \prod_{\substack{s_{i}^\prime\in S_i\setminus\{s_i\}}}\left(s_i-s_i^{\prime}\right)
\right)^{-1}$ for every $\bm{s}=\left(s_1,\ldots,s_m\right)\in \mathcal{S}.$
\end{Def}
Observe that the polynomial $F(\bm{x})$ can be found using interpolation:
\[
F(\bm{x})=\sum_{\left(s_1,\ldots,s_m\right)\in \mathcal{S}}
\frac{
\prod_{i=1}^m \prod_{\substack{s_{i}^\prime\in S_i\setminus\{s_i\}}}\left(x_i-s_i^{\prime}\right)
}
{\left(
\prod_{i=1}^m \prod_{\substack{s_{i}^\prime\in S_i\setminus\{s_i\}}}\left(s_i-s_i^{\prime}\right)
\right)^{2}}.
\]
\begin{Th}\label{20.07.24}
Let $\cS=S_1\times\cdots\times S_m\subseteq K^m$ and $B=\{0,\ldots, n_1-1\}\times \cdots \times \{0,\ldots, n_m-1\}\subset\Z^m$.
Let $F(\bm{x})$ be as defined in Definition~\ref{19.07.24}.
For any $A \subseteq B$, the set $\left\{ \ev_{\cS}(FQ_{\bm{b}}) : \bm{b}\in B\setminus A\right\}$ forms a basis for the dual
$C(\mathcal{S},A)^{\perp}$ of the monomial-Cartesian code $C(\mathcal{S},A).$
\end{Th}
\begin{pf}
Because the definition of $F(\bm{x})$ and $\Res_{\cal S} Q_{\bm{b}},$ it is clear that
$\Res_{\cal S} Q_{\bm{b}}=\ev_{\cS}(FQ_{\bm{b}}).$
\end{pf}
\begin{Lemma}\label{21.07.24}
Let $f_1,\ldots,f_k,g_1,\ldots,g_\ell \in  \cL(A).$
Then $\spn_K\{ \ev_{\cS}(f_1),\ldots,\ev_{\cS}(f_k)\}\subset \spn_K\{ \ev_{\cS}(g_1),\ldots,\ev_{\cS}(g_\ell)\}$ if and only if
$\spn_K\{ f_1,\ldots,f_k\}\subset \spn_K\{ g_1,\ldots,g_\ell\}.$
\end{Lemma}
\begin{pf}
This is a consequence to the fact that the evaluation function $\ev_{\cS}$ is injective.
\end{pf}
Using previous result we can give conditions to determine if a monomial-Cartesian code is self-orthogonal or LCD. An important application
for LCD codes can be found in \cite{Carlet}.
\begin{Th}\label{22.07.24}
Let $\cS=S_1\times\cdots\times S_m\subseteq K^m$ and $A \subseteq B=\{0,\ldots, n_1-1\}\times \cdots \times \{0,\ldots, n_m-1\}\subset\Z^m$.
Let $F(\bm{x})$ be as defined in Definition~\ref{19.07.24}. Then \newline
\noindent
{\rm (a)} $C(\mathcal{S},A)^{\perp}\subset C(\mathcal{S},A)$ if and only if
$\spn_K\left\{\overline{FQ_{\bm{b}}} : \bm{b}\in B\setminus A\right\}\subset \spn_K\left\{\bm{x}^{\bm{a}} : \bm{a}\in A\right\}.$
\noindent
{\rm (b)} $C(\mathcal{S},A)$ is LCD if and only if
$\spn_K\left\{\overline{FQ_{\bm{b}}} : \bm{b}\in B\setminus A\right\}\bigcap \spn_K\left\{\bm{x}^{\bm{a}} : \bm{a}\in A\right\}=0.$
Where $\overline{FQ_{\bm{b}}}$ denotes the normal form of the polynomial $FQ_{\bm{b}}$ with
respect to the Gr\"obner basis $\left\{L_1(x_1),\ldots,L_m(x_m)\right\}.$
\end{Th}
\begin{pf}
The result is a consequence of Lemma~\ref{21.07.24} and Theorem~\ref{20.07.24}.
\end{pf}


Next, we describe some properties for the polynomial $F(\bm{x})$ in order to find conditions that satisfy part (a) from
Theorem~\ref{22.07.24}.
\begin{Prop}
If $q>n_i\geq q/2$ for all $i\in[m],$ then $\deg_{x_i}(F(\bm{x}))\leq q-n_i.$
\end{Prop}\label{19.07.25}
\begin{pf}
Define
\[
F^\prime(\bm{x}):=
\frac{
\prod_{i=1}^m \prod_{\substack{s_{i}^\prime\in K\setminus S_i}}\left(x_i-s_i^{\prime}\right)
}
{\left(-1\right)^{m}}.
\]
Observe that if $\bm{s}=\left(s_1,\ldots,s_m\right)\in \mathcal{S},$ then 
\[
F^\prime(\bm{s})=
\frac{
\prod_{\substack{s_{i}^\prime\in K\setminus S_i}}\left(s_1-s_i^{\prime}\right)
}
{-1}\cdots
\frac{
\prod_{\substack{s_{i}^\prime\in K\setminus S_i}}\left(s_m-s_i^{\prime}\right)
}
{-1}=
\left(
\prod_{i=1}^m \prod_{\substack{s_{i}^\prime\in S_i\setminus\{s_i\}}}\left(s_i-s_i^{\prime}\right)
\right)^{-1}.
\]
Last equality is true because for every $i\in[m]$ we have
$-1=\prod_{\substack{s_{i}^\prime\in K\setminus\{s_i\}}}\left(s_i-s_i^{\prime}\right).$
If $n_i>q/2,$ then $\deg_{x_i}F^\prime(\bm{x})=q-n_i<n_i.$ Thus $F(\bm{x})=F^\prime(\bm{x}),$ because
$F(\bm{x})-F^\prime(\bm{x})\in I(\mathcal{S}).$ If $n_i=q/2,$ then defining $F$ by interpolation we get $\deg_{x_i} F < n_i=q-n_i.$
\end{pf}
The following theorems gives a path to construct quantum and MDS quantum codes.
\begin{Th}\label{19.07.26}
Let $\cS=S_1\times\cdots\times S_m\subseteq K^m$ such that $q>n_i=|S_i|\geq q/2$ for all $i\in[m].$ For
every $\bm{t}=(t_1,\ldots,t_m)\in \{0,\ldots, n_1-\lceil{\frac{q}{2}}\rceil\}\times \cdots \times \{0,\ldots, n_m-\lceil{\frac{q}{2}}\rceil\}\subset\Z^m$
define the set $A_{\bm{t}}=\{0,\ldots, n_1-1-t_1\}\times \cdots \times \{0,\ldots, n_m-1-t_m\},$ then
$C(\mathcal{S},A_{\bm{t}})^{\perp}\subset C(\mathcal{S},A_{\bm{t}}).$
\end{Th}
\begin{pf}
Define $B=\{0,\ldots, n_1-1\}\times \cdots \times \{0,\ldots, n_m-1\}\subset\Z^m$ and take $\bm{b}\in B\setminus A_{\bm{t}}.$
By Theorem~\ref{22.07.24} {\rm (a)} we just need to check that $\overline{FQ_{\bm{b}}} \in \spn_K\left\{\bm{x}^{\bm{a}} : \bm{a}\in A_{\bm{t}}\right\}.$
By definition $Q_{\bm{b}}(\bm{x})=\prod_{i=1}^m q_{i,b_i}(x_i),$ where $L_i(x_i)=x_i^{b_i+1}q_{i,b_i}(x_i)+r_{i,b_i}(x_i).$ It means
$q_{i,b_i}(x_i)\in \spn_K\left\{1,\ldots, x_i^{n_i-b_i-2} \right\}.$ As $\bm{b}\in B\setminus A_{\bm{t}},$ then $n_i-1-t_i<b_i,$ thus $n_i-b_i-2<t_i-1.$
We obtain $\deg q_{i,b_i}(x_i)< t_i-1.$ By Proposition~\ref{19.07.25} $\deg_{x_i}(F(\bm{x}))\leq q-n_i.$
Thus $\deg_{x_i}FQ_{\bm{b}}< q-n_i+t_i-1\leq n_i-1-t_i.$
Last inequality holds because $t_i\leq n_i-\lceil{\frac{q}{2}}\rceil$ and implies
that $\overline{FQ_{\bm{b}}}=FQ_{\bm{b}}\in \spn_K\left\{\bm{x}^{\bm{a}} : \bm{a}\in A_{\bm{t}}\right\}.$
\end{pf}
Now we state an important result to construct stabilizer codes.
\begin{Lemma}\label{19.07.27}\cite[Lemma 17]{AKS}
If there exists a classical linear $[n,k,d]_q$ code $C$ such that $C^{\perp}\subset C$,
then there exists an $[[n, 2k-n, d]]_q$ stabilizer code that is
pure to $d.$ If the minimum distance of $C^{\perp}$ exceeds $d,$ then the
stabilizer code is pure and has minimum distance $d.$
\end{Lemma}
\begin{Th}\label{19.07.28}
Let $\cS=S_1\times\cdots\times S_m\subseteq K^m$ such that $q>n_i=|S_i|\geq q/2$ for all $i\in[m].$ For
every $\bm{t}=(t_1,\ldots,t_m)\in \{0,\ldots, n_1-\lceil{\frac{q}{2}}\rceil\}\times \cdots \times \{0,\ldots, n_m-\lceil{\frac{q}{2}}\rceil\}\subset\Z^m$
there exists an $[[\prod_{i=1}^{m}n_i, 2\prod_{i=1}^{m}(n_i-t_i)-n, \prod_{i=1}^{m}(t_i+1)]]_q$ stabilizer code that is pure to $t_1\cdots t_m.$
\end{Th}
\begin{pf}
The idea is to apply Lemma~\ref{19.07.27} to Theorem~\ref{19.07.26}. By Theorem~\ref{19.07.26}
we have that for $A_{\bm{t}}=\{0,\ldots, n_1-1-t_1\}\times \cdots \times \{0,\ldots, n_m-1-t_m\},$
$C(\mathcal{S},A_{\bm{t}})^{\perp}\subset C(\mathcal{S},A_{\bm{t}}).$ It is clear that the length and dimension of
$C(\mathcal{S},A_{\bm{t}})$ are given by $n$ and $\prod_{i=1}^{m}(n_i-t_i),$ respectively. Finally, the minimum distance comes from Proposition~\ref{19.07.30}.
\end{pf}
Previous result gives a very simple path to prove the existence of quantum error correcting codes with certain parameters.
\begin{Ex}
Let $K=\mathbb{F}_{49}$ and take $n_1=35, n_2=40, t_1=5$ and $t_2=8.$ By Theorem~\ref{19.07.28} we have that there exist the following
quantum error correcting codes: $\displaystyle [[35,25,6]]_{49}$, $\displaystyle [[40,24,9]]_{49}$ and $\displaystyle [[1400,320,54]]_{49}.$
\end{Ex}
Observe that the first two of the previous examples are quantum MDS codes. Actually it is possible to prove the existence of more of them.
\begin{Cor}
For every $q>n\geq q/2$ and every $0\leq t \leq n- \lceil{\frac{q}{2}}\rceil$ there exists an MDS quantum code $[[n,n-2t,t+1]]_q.$
\end{Cor}
\begin{pf}
This is the particular case of Theorem~\ref{19.07.28} when $m=1.$
\end{pf}
\section{Local properties of direct products}\label{19.07.2}
Local properties for linear codes have been studied extensively in the context of distributed storage.
The idea is that every coordinate of a linear code can be used to save the information of a server, so
$n$ servers store a linear code of length $n.$ Informally speaking, a linear code is said to have locality $r$ if for all elements of the code, every coordinate $i$ is a function of other $r$ coordinates. It is
important to remark that the set of these $r$ coordinates depend of $i,$ but not of the codeword.
In terms of distributed storage, locality $r$ means that if one of the $n$ servers fails, then the information of the failed server can be recovered by accessing 
$r$ other servers (rather than $n-1$). If one of these $r$ servers also fails, local recovery might not be possible. For that reason it is useful to have availability. A linear code with availability $t$ means that every coordinate
can be recovered from $t$ pairwise disjoint sets. Formal definitions follow.

\begin{Def}
A linear code $C$ of length $n$ over $K$ is a {\it locally recoverable code} with locality $r$ if
for every position $i\in[n]$ there exist a set $\mathcal{R}_i\subseteq [n]\setminus\{i\}$ and a function
$\phi_i: K^r \to K$ such that $|\mathcal{R}_i|  = r$ and for all $c=(c_1,\ldots, c_n)$ in $C,$
$c_i=\phi_i(c\mid_{\mathcal{R}_i}).$ This definition represents that 
every coordinate $c_i$ for any codeword $c$ can be recovered by the coordinates $c_j,$ where $j\in \mathcal{R}_i.$
The set $\mathcal{R}_i$ is called a {\it recovery set} for the $i$-th position.
\end{Def}
\begin{Def}
A linear code $C$ is said to have $t${\it -availability} with locality $\left(r_1,\ldots, r_t \right)$
if every position $i\in[n]$ has $t$ pairwise disjoint recovery sets $\mathcal{R}_{i1},\ldots, \mathcal{R}_{it}$
with $|\mathcal{R}_{ij}|=r_j,$ for $j\in [t].$
\end{Def}

\begin{Lemma}\label{05.25.19}
Let $C(\cS_1,A)$ and $C(\cS_2,B)$ be locally recoverable monomial-Cartesian codes with localities
$r_1$ and $r_2,$ respectively. The direct product $C(\cS_1\times \cS_2,A\times B)$ has
$2$-availability with locality $(r_1,r_2).$
\end{Lemma}
\begin{pf}
Observe that the coordinates of a monomial-Cartesian code are indexed by the elements of the Cartesian
product, for this reason every position will be given in terms of the elements of the Cartesian product.
Let $\bm{s}_1$ and $\bm{s}_2$ be elements of $\cS_1$ and $\cS_2,$ respectively.
Let $\mathcal{R}_{\bm{s}_1}$ be a recovery set for $\bm{s}_1$ of cardinality
$r_1$ and $\mathcal{R}_{\bm{s}_2}$ a recovery set for $\bm{s}_2$ of cardinality $r_2,$
which exist because
$C(\cS_1,A)$ and $C(\cS_2,B)$ are locally recoverable monomial-Cartesian codes with locality
$r_1$ and $r_2,$ respectively. In the code $C(\cS_1\times \cS_2,A\times B),$
we claim the position $(\bm{s}_1,\bm{s}_2)$ has recovery sets
$\mathcal{R}_{\bm{s}_1}\times \{\bm{s}_2\}$ and $\{\bm{s}_1\}\times \mathcal{R}_{\bm{s}_2}.$

Let $c$ be an element of $C(\cS_1\times \cS_2,A\times B).$ By definition of the direct product, there is a 
polynomial $f(\bm{x},\bm{y})\in \cL(A\times B)\subset K[x_1,\dots, x_{m_1},y_1,\dots,y_{m_2}]$
such that $c=\left(f\left(\bm{s},\bm{s}^\prime\right)
\right)|_{\left(\bm{s},\bm{s}^\prime\right)\in \cS_1\times \cS_2}.$
As $f(\bm{x},\bm{s}_2) \in  \cL(A)\subset K[x_1,\dots,x_{m_1}],$ we can use the set
$\{f(\bm{s},\bm{s}_2) \mid  \bm{s}\in \mathcal{R}_{\bm{s}_1} \}$ to recover the value
$f(\bm{s}_1,\bm{s}_2).$ Thus $\mathcal{R}_{\bm{s}_1}\times \{\bm{s}_2\}$ is a recovery set
for $(\bm{s}_1,\bm{s}_2).$ In analogous way, $\{\bm{s}_1\}\times \mathcal{R}_{\bm{s}_2}$
is a second recovery set for the same position $(\bm{s}_1,\bm{s}_2).$
\end{pf}

We come to the main result of this section, which shows how locally recoverable monomial-Cartesian codes give rise to codes with availability.
\begin{Th}\label{05.26.19}
Let $C(\cS_1,A_1),\ldots, C(\cS_t,A_t)$ be locally recoverable monomial-Cartesian codes with localities
$r_1,\ldots,r_t,$ respectively. The direct product
$C(\cS_1\times \cdots \times \cS_t,A_1\times \cdots \times A_t)$ has
$t$-availability with locality $(r_1,\ldots,r_t).$
\end{Th}
\begin{pf}
This is a consequence of Lemma~\ref{05.25.19} because the product of two monomial-Cartesian
codes is again a monomial-Cartesian code.
\end{pf}
\begin{Rem}
As a corollary of Theorem~\ref{05.26.19} we obtain the family of codes obtained on
\cite[Construction 4]{TamoBarg}, which are direct products of sub-codes of Reed-Solomon codes.
\end{Rem}


\begin{thebibliography}{9}

\bibitem{AKS}{S. A. Aly, A. Klappenecker, P. K. Sarvepalli,
On quantum and classical BCH codes,
IEEE Trans. Inf. Theory {\bf 53} (2007), no. 3,  1183--1188.}



\bibitem{BeelenDatta} P. Beelen, M. Datta,
Generalized Hamming Weights of affine Cartesian Codes,
Finite Fields and Their Applications {\bf 51} (2018) 130--145.

\bibitem{BrME} M. Bras-Amor\'os and M. E. O'Sullivan,
Duality for some families of correction capability optimized evaluation codes,
Adv. Math. Commun {\bf 2} (2008), no.1, 15--33.

\bibitem{BulPel1} S. Bulygin, R. Pellikaan,
Bounded distance decoding of linear error-correcting codes with Gr\"obner bases. 
Journal of Symbolic Computation vol. 44, pp. 1626-1643, 2009. 

\bibitem{BulPel2} S. Bulygin, R. Pellikaan,
Decoding and finding the minimum distance with Gr\"obner bases : history and new insights. 
in Series on Coding Theory and Cryptology vol. 7 
Selected Topics in Information and Coding Theory, 
I. Woungang, S. Misra and S.C. Misra, Eds., vol. 7, pp. 585--622, World Scientific 2010.

\bibitem{BulPel3} S. Bulygin, R. Pellikaan,
Decoding error-correcting codes with Gr\"obner bases,. 
Proceedings of the 28-th Symposium on Information Theory in the Benelux, WIC 2007, 
R. Veldhuis, H. Cronie, H. Hoeksema, Eds., Enschede, May 24-25, pp. 3-10, 2007. 

\bibitem{BulPel4} S. Bulygin, R. Pellikaan,
Decoding linear error-correcting codes up to half the minimum distance with Gr\"obner bases. 
In Gr\"obner Bases, Coding, and Cryptography, M. Sala, T. Mora, L. Perret, S. Sakata, and C. Traverso Eds., 
pp. 361-365, Springer, Berlin, 2009. 
 
\bibitem{CS}
A. R. Calderbank, P. W. Shor, Good quantum error-correcting codes
exist, Phys. Rev. A 54 (1996), 1098--1105.

 
\bibitem{Carlet}
C. Carlet and S. Guilley, 
Complementary Dual Codes
for Counter-Measures to Side-Channel Attacks, Advances in Mathematics of Communications {\bf 10} (2016) 131--150.

 \bibitem{Carlet_equiv}
 C. Carlet, S. Mesnager, C. Tang, Y. Qi, and R. Pellikaan,  Linear codes over $\F_q$ are equivalent to LCD codes for $q > 3$, IEEE Transactions on Information Theory {\bf 64} (2018), no. 4, 3010--3017.
 
 
\bibitem{carvalho} C. Carvalho, 
On the second Hamming weight of some Reed-Muller type codes,
Finite Fields and Their Applications {\bf 24} (2013) 88--94.

\bibitem{carvalho2} C. Carvalho, V. G. Neumann,
On the next-to-minimal weight of affine Cartesian codes,
Finite Fields and Their Applications {\bf 44} (2017) 113--134.

\bibitem{carvalho3} C. Carvalho, V. G. Neumann,
Projective Reed–Muller type codes on rational normal scrolls,
Finite Fields and Their Applications {\bf 37} (2016) 85--107.

\bibitem{carvalho4} C. Carvalho, V. G. Neumann, H. H. L\'opez,
Projective Nested Cartesian Codes,
Bull Braz Math Soc, New Series (2016).


\bibitem{CLO1}{D. Cox, J. Little and D. O'Shea,
{\it Ideals, Varieties, and Algorithms\/},
Undergraduate Texts in Mathematics, Springer-Verlag, Third Edition, 2008.}


\bibitem{Eisen}{D. Eisenbud, {\it Commutative Algebra with a view
toward Algebraic Geometry\/}, Graduate
Texts in  Mathematics {\bf 150}, Springer-Verlag, 1995.}


\bibitem{FaSh}{J. Farr and S. Gao,
Gr\"obner bases, Pad\'e  approximation, and decoding of linear codes,
Coding Theory and Quantum Computing, Contemporary Mathematics, Amer. Math. Soc., Providence, RI, {\bf 381} (2005), 3--18.}


\bibitem{GOlav}
C. Galindo, O. Geil, F. Hernando, D. Ruano, 
On the distance of stabilizer quantum codes from J-affine variety codes,
Quantum Inf Process (2017) 16: 111. https://doi.org/10.1007/s11128-017-1559-1

\bibitem{GHRuano}
C. Galindo, F. Hernando, D. Ruano,
Stabilizer quantum codes from J-affine variety codes and a new Steane-like enlargement,
Quantum Inf Process (2015) 14: 3211. https://doi.org/10.1007/s11128-015-1057-2

\bibitem{Geil} O. Geil, C. Thomsen,
Weighted Reed-Muller codes revisited,
Des.\ Codes Cryptogr. \textbf{66}(1--3) (2013) 195--220.

\bibitem{GRT} M. Gonz\'alez-Sarabia, C. Renter\'\i a and H.
Tapia-Recillas, Reed-Muller-type codes over the Segre variety,  
Finite Fields Appl. {\bf 8}  (2002),  no. 4, 511--518.

\bibitem{harris} J. Harris, {\it Algebraic Geometry. A first course},

\bibitem{Han1} J. P. Hansen,
Toric Surfaces and Error-correcting Codes,
Coding Theory, Cryptography and Related Areas,  Springer, Berlin, Heidelberg, 132--142.

\bibitem{huf-pless} W. Huffman and V. Pless,
{\it Fundamentals of Error-Correcting Codes},
Cambridge University Press, Cambridge, 2003.

\bibitem{joyner-decoding} D. Joyner, Toric codes over finite fields, 
Appl. Algebra Engrg. Comm. Comput. {\bf 15} (2004), no. 1, 63--79.

\bibitem{KKKK}{A. Ketkar, A. Klappenecker, S. Kumar, P. K. Sarvepalli,
Nonbinary Stabilizer Codes over Finite Fields,
IEEE Transactions on Information Theory {\bf 52} (2006), no. 11, 4892--4914.}


\bibitem{lopez-manga-matt} H. H. L\'opez, F. Manganiello, G. L. Matthews,
Affine Cartesian codes with complementary duals,
Finite Fields and Their Applications, {\bf 57} (2019), 13--28.

\bibitem{lopez-villa} H. H. L\'opez, C. Renter\'ia-M\'arquez, R. H. Villarreal,
Affine Cartesian codes,
Des.\ Codes Cryptogr. \textbf{71}(1) (2014) 5--19.

\bibitem{MacWilliams-Sloane} F. J. MacWilliams and N. J. A. Sloane, 
The Theory of Error-correcting Codes, North-Holland, 1977.

\bibitem{JMassey} James L. Massey,
Linear codes with complementary duals,
Discrete Mathematics, {\bf 106--107} (1992), 337--342.


\bibitem{renteria-tapia-ca2} C. Renter\'\i a and H. Tapia-Recillas, 
Reed-Muller codes: an ideal theory approach, Comm. Algebra {\bf 25}
(1997), no. 2, 401--413.

\bibitem{Mun}{C. Munuera,
Locally Recoverable Codes with Local Error Detection,
\url{https://arxiv.org/pdf/1812.00834.pdf}}

\bibitem{DRua}{D. Ruano,
On the structure of generalized toric codes,
J. Symbolic Comput., {\bf 44} (2009), no. 5, 499--506.}

\bibitem{Sop2} I.~Soprunov, J.~Soprunova, { Toric surface codes and Minkowski length of polygons}, SIAM J. Discrete Math. {\bf 23}, Issue 1, (2009) 384--400 

\bibitem{Sop}{I. Soprunov, J. Soprunova,
Bringing Toric Codes to the Next Dimension,
SIAM Journal on Discrete Mathematics, {\bf 24} (2010), no. 2, 655--665.
}

\bibitem{Sop3}{I. Soprunov, Lattice polytopes in coding theory,
Journal of Algebra Combinatorics Discrete Structures and Applications (2015), Vol 2, No 2, 85--94
}




\bibitem{Steane}
A. M. Steane, Multiple-particle interference and quantum error correction.
Proc. Roy. Soc. London Ser. A 452 (1996), no. 1954, 2551--2577.

\bibitem{TamoBarg} I. Tamo, A. Barg,
A Family of Optimal Locally Recoverable Codes,
IEEE Transactions on Information Theory {\bf 60} (2014), no. 8, 4661--4676.



\bibitem{tsfasman} M. Tsfasman, S. Vladut and D. Nogin, {\it
Algebraic 
geometric codes{\rm:} basic notions}, Mathematical Surveys and
Monographs {\bf 139}, American Mathematical Society, 
Providence, RI, 2007. 

\bibitem{van-lint} J. H. van Lint, {\it Introduction to coding theory}, Third
edition, Graduate Texts in Mathematics {\bf 86}, Springer-Verlag,
Berlin, 1999. 

\bibitem{monalg}{R. H. Villarreal, {\it Monomial Algebras\/},
second edition, Monographs and 
and Research notes in Mathematics, 2015.} 


\bibitem{Wei}{V. K. Wei, K. Yang,
On the Generalized Hamming Weights of Product Codes,
IEEE Transactions on Information Theory, {\bf 30} (1993), no. 5, 1079-1713.
}

\end{thebibliography}
\end{document}